\documentclass[twocolumn]{jpsj2}

\title{Thermohydrodynamics in Quantum Hall Systems}

\author{Hiroshi \textsc{Akera}\thanks{E-mail address: akera@eng.hokudai.ac.jp} 
and Hidekatsu \textsc{Suzuura}\thanks{E-mail address: suzuura@eng.hokudai.ac.jp}}

\inst{Department of Applied Physics, Graduate School of Engineering, 
Hokkaido University, Sapporo 060-8628}

\recdate{\today}

\abst{
A theory of thermohydrodynamics 
in two-dimensional electron systems in quantizing magnetic fields 
is developed including a nonlinear transport regime. 
Spatio-temporal variations of 
the electron temperature and the chemical potential in the local equilibrium
are described by the equations of conservation 
with the number and thermal-energy flux densities.   
A model of these flux densities due to hopping and drift processes 
is introduced for a random potential varying slowly compared to 
both the magnetic length and the phase coherence length. 
The flux measured in the standard transport experiment is derived 
and is used to define a transport component of the flux density. 
The equations of conservation can be written in terms of 
the transport component only. 
As an illustration, the theory is applied to the Ettingshausen effect, 
in which a one-dimensional spatial variation 
of the electron temperature is produced perpendicular to the current.  
}

\kword{integer quantum Hall effect, Ettingshausen effect, 
nonlinear transport, local equilibrium, thermohydrodynamics, theory}

\begin{document}
\maketitle

\def\muec{\mu_{\rm ec}}
\def\much{\mu}
\def\lvh{l_{\rm vh}}
\def\vj{\mib j}
\def\vJ{\mib J}
\def\vI{\mib I}
\def\vb{\mib b}
\def\vB{\mib B}
\def\vM{\mib M}
\def\vE{\mib E}
\def\vF{\mib F}
\def\vr{\mib r}
\def\vn{\mib n}
\def\vt{\mib t}
\def\vnabla{\mib \nabla}

\def\Lnnxx{L^{11}_{xx}}
\def\Lnqxx{L^{12}_{xx}}
\def\Lqnxx{L^{21}_{xx}}
\def\Lqqxx{L^{22}_{xx}}

\def\Lnnxy{L^{11}_{xy}}
\def\Lnqxy{L^{12}_{xy}}
\def\Lqnxy{L^{21}_{xy}}
\def\Lqqxy{L^{22}_{xy}}

\def\Lnnyx{L^{11}_{yx}}
\def\Lnqyx{L^{12}_{yx}}
\def\Lqnyx{L^{21}_{yx}}
\def\Lqqyx{L^{22}_{yx}}

\def\Lnnyy{L^{11}_{yy}}
\def\Lnqyy{L^{12}_{yy}}
\def\Lqnyy{L^{21}_{yy}}
\def\Lqqyy{L^{22}_{yy}}

\def\K21yx{K^{21}_{yx}}
\def\Aqnyx{A^{21}_{yx}}
\def\Aqqyy{A^{22}_{yy}}
\def\Bqnyy{B^{21}_{yy}}
\def\Bqqyy{B^{22}_{yy}}

\def\sxx{\sigma_{xx}}
\def\sxy{\sigma_{xy}}
\def\syx{\sigma_{yx}}
\def\syy{\sigma_{yy}}
\def\rxx{\rho_{xx}}
\def\rxy{\rho_{xy}}
\def\ryx{\rho_{yx}}
\def\ryy{\rho_{yy}}
\def\kB{k_{\rm B}}
\def\Te{T_{\rm e}}
\def\TL{T_{\rm L}}
\def\PL{P_{\rm L}}
\def\oc{\omega_{\rm c}}
\def\ve{\varepsilon}
\def\Cp{C_{\rm p}}
\def\edge{{\rm edge}}
\def\bulk{{\rm bulk}}
\def\tr{\rm tr}
\def\Streda{St$\check{\rm r}$eda}
%

\section{Introduction} 

Studies on spatial and temporal variations of 
thermodynamic and hydrodynamic variables 
reveal important properties of the studied 
system.~\cite{Groot1962,Scholl2001} 
A length scale and a time scale of variations, 
which occur in response to an external perturbation 
as well as in self-organization, 
are fundamental quantities 
which are never obtained by 
studying the system in uniform steady states. 

Studies in this direction have already been performed by many authors 
in two-dimensional electron systems in quantizing magnetic fields, 
which exhibit the quantum Hall effect in the region of low temperatures 
and small currents.~\cite{Klitzing1980,Kawaji1981}
Theoretical\cite{Chklovskii1992} and experimental
studies on spatial variations of the electrostatic potential in equilibrium 
have revealed a unique spatial pattern consisting of  
compressible and incompressible strips, 
which is a manifestation of the Landau quantization. 
These studies have recently been extended to 
nonequilibrium states with the applied current,  
and the response of the electrostatic potential to the applied current 
has been investigated experimentally\cite{Ahlswede2001} 
and theoretically.\cite{Guven2003}
However, most of these previous works have been restricted to  
quantities related to 
the chemical potential and the electrostatic potential.  
The electron temperature is known to be the key variable 
in the large-current regime where 
the breakdown of the quantum Hall 
effect~\cite{Ebert1983,Cage1983,Kuchar1984} takes place.
The spin density is the order parameter of the ferromagnetic phase 
which appears when the Landau level filling factor is an odd integer.  
For a complete description of 
thermohydrodynamics in quantum Hall systems,  
studies need to be extended to 
spatial variations of such variables and to their temporal variations.

Spatio-temporal variations of the electron temperature 
have been investigated indirectly by measuring 
spatial~\cite{Komiyama1996,Kaya1998,Kaya1999} 
and temporal~\cite{Cage1983,Ahlers1993,Sagol2001} variations
of the diagonal component of the resistivity tensor. 
Kawano and Komiyama~\cite{Kawano2000,Kawano2003} 
have recently investigated the electron temperature 
more directly by measuring the local cylotron emission intensity 
using a scanning optical probe,   
and have revealed a variety of spatial variations of the electron temperature.  
The first attempt to construct a general theory 
for spatio-temporal variations of the electron temperature and the chemical potential 
in quantum Hall systems 
was made by one of the present authors.~\cite{Akera2002} 
However, this previous theory has been found 
to employ an inexact formula of the thermal flux density. 

In this paper we develop a theory of 
thermohydrodynamics in quantum Hall systems, 
which is a revision and a generalization of the previous theory.~\cite{Akera2002}   
We describe spatio-temporal variations of 
the electron temperature and the chemical potential
in the local equilibrium including the nonlinear transport regime 
with use of the equations of conservation, 
as in the previous theory.~\cite{Akera2002} 
A revision is made by writing the equations in terms of 
the total thermal-energy flux density due to the drift motion.   
We generalize the previous theory 
to study thermohydrodynamics 
in the presence of a nonuniform potential 
such as a slowly-varying confining potential. 

The organization of the paper is as follows. 
In \S 2, 
we introduce a model, assumptions, and macroscopic variables.  
Random potential due to ionized impurities is assumed to be 
slowly varying in the scale of 
both the magnetic length and the phase coherence length. 
In \S 3, 
we describe the equations of conservation and 
introduce a model of the number and thermal-energy flux densities 
due to hopping and drift processes.  
Each of the flux densities is a linear function of 
gradients of the electron temperature, the electrochemical potential, 
and the electrostatic potential.  
The coefficients in these linear functions depend on 
the local electron temperature and the local chemical potential. 
Such dependences are the origin of nonlinearity in this theory. 
In \S 4, 
we derive formulas for fluxes measured 
in the standard transport experiment 
for a narrow sample with a steep confining potential. 
In \S 5, 
we define a transport flux density for each of the measured fluxes. 
We show that the transport flux density defined in this manner 
is equal to the total flux density minus the magnetization current density. 
We find that the equations of conservation can be written in terms of 
the transport component only. 
It is shown that the transport flux densities  
satisfy the Einstein relation and the Onsager relation,  
while the total flux densities do not. 
In \S 6, as an illustration, we apply our theory to the Ettingshausen effect, 
in which a one-dimensional spatial variation 
of the electron temperature is produced perpendicular to the current. 
In \S 7, we review briefly the existing literature 
on the thermoelectric-transport theory in quantizing magnetic fields
and on thermohydrodynamics in quantum Hall systems.   
In \S 8, conclusions are given. 

\section{Model, Assumptions and Macroscopic Variables}

\subsection{Drift and Hopping Processes}  

We consider a two-dimensional electron system 
(in the $xy$ plane)  
in a perpendicular magnetic field $\vB=(0,0,B)$.  
The local potential energy $V_{\rm loc}$ 
consists of 
the disorder potential due to ionized donors 
with the fluctuation length scale 
of $\lvh \sim 0.1\ {\rm \mu m}$~\cite{Nixon1990} 
and the macroscopic potential $V$.  
The macroscopic potential $V$ consists of 
the potential in equilibrium such as the confining potential 
and that induced by the applied current.  
The fluctuation length scale $\lvh$ 
is much longer than 
the magnetic length $l=\sqrt{\hbar c /e|B|}$ ($e>0$)
which is about $0.01\ {\rm \mu m}$ at $|B|=5$T. 

An electron loses the phase coherence by  
a scattering from other electrons and phonons. 
The phase coherence length is assumed to be less than $\lvh$. 
Therefore the Anderson localization due to the interference in the disorder potential 
as well as the energy quantization for closed orbits around potential hills and valleys 
are negligible. 
Then we employ the classical-drift model~\cite{Iordansky1982,Prange1982,Apenko1985}, 
in which each electron state is described by a wave packet with extent of the order of $l$ 
drifting along the equipotential line. 
The energy of the wave packet at $\vr_{\rm wp}$ 
in the Landau level specified by a set of quantum numbers $\alpha=\{N, \sigma\}$ 
with the Landau index $N=0,1,2,\cdots$ and spin $\sigma$ is given by 
\begin{equation}
\ve^0_{\alpha} + V_{\rm loc}(\vr_{\rm wp}) ,
\end{equation}
where $\ve^0_{\alpha}$ is the energy of the Landau level
in the absence of the potential.
The wave packet  follows 
the equipotential line of $V_{\rm loc}$ with a velocity given by 
\begin{equation}
\frac{d \vr_{\rm wp}}{d t}
= \frac{l^2}{\hbar} s_B \hat{\epsilon} \vnabla V_{\rm loc}(\vr_{\rm wp}) , \ 
s_B=\frac{B}{|B|}, \
\hat{\epsilon}= \left( \!\! \begin{array}{cc}
                                  0 & 1 \\
                                 -1 & 0 
            \end{array}  \!\! \right)  .	 
\label{eq:Xt}
\end{equation}

A wave packet with energy at the center of the broadened Landau level 
follows the extended equipotential line of the disorder potential  
and contributes to the macroscopic current by 
the drift motion given by eq.(\ref{eq:Xt}). 
Other wave packets follow closed orbits within each vh-region. 
Here a vh-region is a subspace which is 
bounded by extended equipotential lines 
and contains one potential hill or valley. 
A hopping of such a localized wave packet between neighboring vh-regions 
by a scattering from other electrons 
contributes to the macroscopic current.~\cite{Akera2000}   
A tunneling through a saddle point of the potential 
also produces the macroscopic current, 
although its contribution is estimated to be small.~\cite{Ise2002} 
Compared to such intra-Landau-level transitions, 
the contribution of inter-Landau-level transitions to the macroscopic current 
is negligible in the slowly-varying potential.\cite{Akera2000} 
In this paper we consider the drift and the hopping processes 
within each Landau level in calculating the macroscopic current.

\subsection{Local Equilibrium}  
\label{sec:Local_Equilibrium}

We employ the local-equilibrium approximation, 
which has been used extensively in various systems 
and is described in detail, for example, in refs.\citen{Groot1962} and \citen{Cooper1997}. 
We assume the local equilibrium within each vh-region, which means that   
the energy distribution of electrons in each vh-region is described by 
the Fermi distribution function with 
the electron temperature $\Te$ and the electrochemical potential $\muec$
of the vh-region: 
\begin{equation}
f( \ve, \muec, \Te ) = 1/\{\exp[(\ve -\muec)/\kB \Te] +1\} . 
\end{equation}
Each of $\Te$ and $\muec$ differs from vh-region to vh-region 
in nonequilibrium states. 
The same electron distribution within the condition of constant $\Te$ 
has been used in a theory on the quantum Hall effect.~\cite{Iordansky1982}
To maintain the local equilibrium, 
it is necessary that the applied current and the magnetic field should vary slowly in time 
compared with the time necessary for the relaxation to the local equilibrium. 
The slowest process involved in this relaxation is 
the transition of electrons between Landau levels  
in the case where the distribution of electrons to different Landau levels 
deviates from that in the local equilibrium. 
We also assume that phonons are in equilibrium 
with the lattice temperature $\TL$, 
and that $\TL$ does not change by the presence of the applied current.

\subsection{Spatial Averaging and Macroscopic Variables}  

Since the potential due to ionized donors is random,  
quantities such as electron and energy densities fluctuate 
in the length scale of $\lvh \sim 0.1 \ {\rm \mu m}$. 
We eliminate such random fluctuations by a spatial averaging,  
while we retain variations in the scale larger than $\lvh$ 
which are produced by the confining potential and by the applied current. 
We assume that $\Te$ and $\muec$ in nonequilibrium vary slowly in space 
compared with $\lvh$.
We note that the spatial variation of the measured cyclotron emission intensity  
is also macroscopic because of 
the limited spatial resolution of $50 \ {\rm \mu m}$.~\cite{Kawano2003} 

Thermohydrodynamics studies spatio-temporal variations of 
macroscopic variables. 
The macroscopic variables in the present system are 
the electron temperature $\Te(\vr,t)$, 
the electrochemical potential $\muec(\vr,t)$, 
and the potential energy $V(\vr,t)$. 
The chemical potential $\much(\vr,t)$ is defined by 
$\much = \muec - V$. 
The macroscopic electric field $\vE(\vr,t)$ is given by 
$\vE=\vnabla V /e $.  
The macroscopic potential $V(\vr,t)$ is determined by 
electrostatics in terms of the averaged electron density 
which is a function of $\Te$ and $\much$. 
Other macroscopic variables $\Te$ and $\muec$ 
are determined by the following equations of conservation.

\section{Formulation with the Total Fluxes}

\subsection{Equations with the Total Flux Densities}

There are two equations of conservation in our model of quantum Hall systems. 
One describes the conservation of the electron number,  
and the other describes the conservation of the energy. 
The equation of the momentum conservation is absent 
since in the classical-drift model 
the velocity of the wave packet is not an additional degree of freedom but 
given in eq.(\ref{eq:Xt}) to be a function of its position. 

The equation of the electron number conservation is 
\begin{equation}
\frac{\partial n}{\partial t} = -\vnabla \cdot \vj_n \ ,
\label{eq:nt}
\end{equation}
where $n$ is the electron density and 
$\vj_n$ is the number flux density.  
The equation of the energy conservation is 
\begin{equation}
\frac{\partial \epsilon}{\partial t} = -\vnabla \cdot \vj_{\epsilon} \ - \PL.
\label{eq:et}
\end{equation}
Here $\epsilon$ is the energy density and   
$\vj_{\epsilon}$ is the energy flux density, while 
$\PL$ is the energy loss per unit area at point $\vr$ 
due to the heat transfer between electrons and phonons 
and is in general a function of $\much$, $\Te$, and $\TL$. 
The energy density 
is the sum of the kinetic energy density and the potential energy density:  
$\epsilon = \epsilon_{\rm kin}+nV$. 

Equations describing the time evolution of $\much$ and $\Te$ 
are derived from those of $n(\much, \Te)$ and the entropy density $s(\much, \Te)$, 
respectively.   
The equation for the time evolution of $s$ is derived 
using eqs. (\ref{eq:nt}), (\ref{eq:et}), and 
\begin{equation}
\Te ds = d\epsilon - \muec dn , 
\label{eq:tds}
\end{equation}
to be 
\begin{equation}
\Te \frac{\partial s}{\partial t} = -\vnabla \cdot \vj_q \ 
                              - \vnabla \muec \cdot   \vj_n   - \PL,
\label{eq:tst}
\end{equation}
where $\vj_q$ is the thermal flux density defined by 
\begin{equation}
\vj_q = \vj_{\epsilon} - \muec \vj_n  . 
\label{eq:jq}
\end{equation}

\subsection{Number and Thermal Flux Densities}

\subsubsection{Hopping Components}

First we consider the number flux between neighboring vh-regions 
due to hopping processes, denoted as $J_n^{\rm hop}$. 
Since we consider only hopping processes within each Landau level, we have 
$J_n^{\rm hop}=\sum_{\alpha} J_{n\alpha}^{\rm hop}$.  
Each $J_{n\alpha}^{\rm hop}$ is induced by 
the difference in the electron temperature $\Delta \Te$ and 
that in the electrochemical potential $\Delta \muec$ between the two vh-regions.
In the first order of $\Delta \Te$ and $\Delta \muec$, 
$J_{n\alpha}^{\rm hop}$ is given by 
\begin{equation}
J_{n\alpha}^{\rm hop} =  A_{\alpha} \Delta \muec +B_{\alpha} \Delta \Te  .  
\end{equation}
In the following we show that the coefficients $A_{\alpha}$ and $B_{\alpha}$ 
are related to each other in our model of hopping processes. 

The wave packets contributing to the hopping processes are only those in the vicinity of 
the boundary between the two vh-regions,  
since the transition rate is negligible when 
the distance between the wave packets is much larger than $l$. 
The energies of such wave packets 
are confined within an energy range around $\ve_{\alpha}(\vr)$ 
with width $\Gamma_{\rm hop}$ 
where $\ve_{\alpha}(\vr) =  \ve^0_{\alpha} + V(\vr)$ 
and $\Gamma_{\rm hop} \sim \Gamma l/\lvh$ 
with $\Gamma$ the width of the broadened Landau level.  
Therefore 
the corresponding occupation probability is well approximated to be 
\begin{equation}
f_{\alpha} =  f( \ve_{\alpha},\muec, \Te ) ,
\end{equation}
when $\Gamma_{\rm hop} \ll \kB\Te$.  
In the case of elastic hopping processes between 
a vh-region with $\muec$ and $\Te$ 
and its neighbor with $\muec+\Delta\muec$ and $\Te+\Delta \Te$, 
$J_{n\alpha}^{\rm hop}$ is given by
\begin{equation}
\!\!\!\!\!J_{n\alpha}^{\rm hop} = 
- C_{\alpha} 
[f( \ve_{\alpha}, \!\muec+\!\Delta\muec, \Te+\!\Delta \Te ) - f( \ve_{\alpha}, \muec, \Te)] . 
\end{equation}
In the first order of $\Delta \Te$ and $\Delta \muec$, we obtain 
\begin{equation}
J_{n\alpha}^{\rm hop} = - C_{\alpha} 
\left( \frac{\partial f_{\alpha}}{\partial \muec}\Delta \muec 
+ \frac{\partial f_{\alpha}}{\partial \Te} \Delta \Te \right). 
\label{eq:Jn_hop}
\end{equation}
Also in the case of inelastic electron-electron scatterings, 
this formula is approximately applicable 
as long as the energy change in each hopping process
is much smaller than $\kB\Te$.
We assume eq.(\ref{eq:Jn_hop}) in the following. 

The hopping number flux density $\vj^{\rm hop}_n$ 
averaged in the macroscopic scale 
is then given by 
\begin{equation}
\vj^{\rm hop}_{n}=  - \sum_{\alpha} D_{\alpha} 
      \left(  \frac{\partial f_{\alpha}}{\partial \muec } \vnabla \muec 
          +\frac{\partial f_{\alpha}}{\partial \Te } \vnabla \Te  \right) .
\label{eq:jn_hop0}
\end{equation}
Here $D_{\alpha}$ is written in terms of the transition rate of each hopping process,    
which depends on the disorder potential. 
The disorder potential in turn is a function of $\much$ and $\Te$ 
since it is screened by electrons.  

We introduce transport coefficients $\Lnnxx$ and $\Lnqxx$ 
and write $\vj^{\rm hop}_{n}$ as 
\begin{equation}
\vj^{\rm hop}_{n}=  - \Lnnxx \vnabla \muec  - \Lnqxx \Te^{-1} \vnabla \Te .
\label{eq:jn_hop}
\end{equation}
From eq.(\ref{eq:jn_hop0}) we have
\begin{eqnarray}
\Lnnxx  =  e^{-2}\sxx =  (\kB\Te)^{-1} \sum_{\alpha} D_{\alpha} 
             f_{\alpha} (1- f_{\alpha}) ,  \\
\Lnqxx  =   (\kB\Te)^{-1} \sum_{\alpha} D_{\alpha} 
  f_{\alpha} (1- f_{\alpha})  ( \ve^0_{\alpha} - \much ) .  
\end{eqnarray}
When we assume the $N$ dependence of $D_{\alpha}$ to be $D_{\alpha}= (2N+1) D_0$ 
with $D_0$ the coefficient for $N=0$, 
the above formula of $\Lnnxx$ coincides 
with that for short-range scatterings 
in the self-consistent Born approximation\cite{Ando1974} 
in the case where $\Gamma \ll \kB \Te$.  
Note that the above formula is applicable 
as long as $\Gamma_{\rm hop} \ll \kB\Te$. 
The coefficients $\Lnnxx$ and $\Lnqxx$ are functions of $\Te(\vr,t)$ and $\much (\vr,t)$. 
In the linear-response regime, 
$\Lnnxx$ and $\Lnqxx$ are to be evaluated in equilibrium  
and eq.(\ref{eq:jn_hop}) reduces to 
the so-called phenomenological equation.~\cite{Groot1962}  
In our model, a nonlinear effect is taken into account 
in dependences of $\Lnnxx$ and $\Lnqxx$ on 
deviations of $\Te(\vr,t)$ and $\much (\vr,t)$ from their equilibrium values. 

The thermal flux density [eq.(\ref{eq:jq})] is given similarly, in hopping processes, by 
\begin{equation}
\vj^{\rm hop}_{q}=  - \sum_{\alpha} (\ve_{\alpha} - \muec ) D_{\alpha} 
      \left(  \frac{\partial f_{\alpha}}{\partial \muec } \vnabla \muec 
          +\frac{\partial f_{\alpha}}{\partial \Te } \vnabla \Te  \right) .
\end{equation}
This expresses that an electron in the Landau level $\alpha$ 
carries a thermal energy $\ve_{\alpha}-\muec$. 
In terms of transport coefficients, $\vj^{\rm hop}_{q}$ is expressed as  
\begin{equation}
\vj^{\rm hop}_{q}=  - \Lqnxx \vnabla \muec  - \Lqqxx \Te^{-1} \vnabla \Te ,
\end{equation}
where 
\begin{eqnarray}
&\Lqnxx =  \Lnqxx  \ , \\
&\Lqqxx = (\kB\Te)^{-1} \sum_{\alpha} D_{\alpha} 
  f_{\alpha} (1- f_{\alpha})  ( \ve^0_{\alpha} - \much )^2 . 
\end{eqnarray}

\subsubsection{Drift Components}

The local flux density due to the drift motion fluctuates spatially since  
the local potential $V_{\rm loc}$ contains the random potential. 
We average the local flux density to obtain the macroscopic flux density. 
The macroscopic number flux density in the Landau level $\alpha$, 
$\vj^{\rm drift}_{n\alpha}$, is written as
\begin{equation}
\vj^{\rm drift}_{n\alpha} = 
    \left\langle f(\ve^0_{\alpha}+V_{\rm loc}, \muec,\Te)   
    h^{-1} s_B \hat{\epsilon} \vnabla V_{\rm loc}
     \right\rangle_{\rm av} . 
\end{equation}
Since localized states make no contributions 
to the macroscopic flux density, 
the occupation probability of localized states in the above equation 
can be replaced by  
that of extended states. Then we have 
\begin{equation}
\vj^{\rm drift}_{n\alpha} =  f(\ve^0_{\alpha}+V, \muec,\Te)
    \left\langle    h^{-1} s_B \hat{\epsilon} \vnabla V_{\rm loc}
    \right\rangle_{\rm av} .  
\end{equation}
Since the spatial average of $\vnabla V_{\rm loc}$ is equal to $\vnabla V$,  
then the number flux density $\vj^{\rm drift}_{n}$ 
due to all the Landau levels becomes 
\begin{equation}
\vj^{\rm drift}_{n} =  \Lnnyx \hat{\epsilon} \vnabla V , 
\label{eq:jn_drift}
\end{equation}
with
\begin{equation}
\Lnnyx = \frac{\syx}{e^2}= \frac{s_B}{h} \sum_{\alpha} f_{\alpha} .
\end{equation}
Similarly we have for the thermal flux density 
\begin{equation}
\vj^{\rm drift}_{q} =  \K21yx \hat{\epsilon} \vnabla V , 
\end{equation}
with
\begin{equation}
\K21yx = \frac{s_B}{h} \sum_{\alpha} (\ve^0_{\alpha} - \much) f_{\alpha} .
\end{equation}

The coefficients $\Lnnyx$ and $\K21yx$ can be written as 
\begin{equation}
\Lnnyx = s_B  \frac{2\pi l^2}{h} n_0 \ , 
\end{equation}
\begin{equation}
\K21yx = s_B \frac{2\pi l^2}{h} (\Omega_0 +\Te s_0 ) \ , 
\label{eq:K21}
\end{equation}
in terms of thermodynamic quantities $n_0$, $\Omega_0$, and $s_0$. 
Here $\Omega_0$ is 
the thermodynamic potential density 
corresponding to the grand canonical ensemble 
in the absence of disorder ($V_{\rm loc}=V$):  
\begin{equation}
\!\!\!\! \Omega_0(\Te,\much,B) \!= \!
    - \frac{\kB \Te}{2\pi l^2}  \sum_{\alpha}
      \ln \! \left[1 \!+\! \exp \! \left( \!- \frac{\ve^0_{\alpha} \!-\! \much}{\kB \Te} \right) 
         \right]  ,
\end{equation}
while $n_0$ and $s_0$ are the electron density and the entropy density, 
respectively, in this case: 
\begin{equation}
n_0= - \left( \frac{\partial \Omega_0}{\partial \much}\right)_{\Te,B} , \ \ \ 
s_0= - \left( \frac{\partial \Omega_0}{\partial \Te}\right)_{\mu,B} . 
\end{equation}


\subsubsection{Total Flux Densities}

The total flux densities 
$\vj_n = \vj^{\rm hop}_{n} + \vj^{\rm drift}_{n}$ and 
$\vj_q = \vj^{\rm hop}_{q} + \vj^{\rm drift}_{q}$ are written as  
\begin{equation}
j_{nx}= - \Lnnxx \nabla_x \muec 
                             + \Lnnyx \nabla_y V 
                             -  \Lnqxx \Te^{-1} \nabla_x \Te , 
\label{eq:jnx}
\end{equation}
\begin{equation}
j_{ny}= - \Lnnyx \nabla_x V 
                             - \Lnnxx \nabla_y \muec 
                             - \Lnqxx \Te^{-1}  \nabla_y \Te , 
\label{eq:jny}
\end{equation}
\begin{equation}
j_{qx}= - \Lnqxx \nabla_x \muec 
                             + \K21yx \nabla_y V 
                             -  \Lqqxx \Te^{-1}  \nabla_x \Te , 
\label{eq:jqx}
\end{equation}
\begin{equation}
j_{qy}= - \K21yx \nabla_x V 
                             - \Lnqxx \nabla_y \muec 
                             - \Lqqxx \Te^{-1}  \nabla_y \Te .  
\label{eq:jqy}
\end{equation}
The total flux densities are produced not only by 
$\vnabla \muec$ and $\vnabla \Te$, but also by $\vnabla V$, 
and therefore they are in general nonzero in equilibrium. 
The standard transport experiment measures 
a flux through a cross section of the sample, which is zero in equilibrium. 
Such a flux, which is induced in nonequilibrium, is calculated in the next section.

\section{Measured Fluxes in a Narrow Wire} 

\subsection{Model and Assumptions}
\label{sec:model}

In this section we derive formulas for the fluxes 
measured in the standard transport experiment. 
We consider a narrow two-dimensional system along the $x$ direction 
and separate the system into the bulk region $0<y<\Delta y$ 
and the edge region $y<0, \Delta y<y$. 
In the edge region we assume the presence of a confining potential which 
increases to infinity 
so that the electron density decreases to zero 
within a length scale of the order of $\lvh$. 
The width of the edge region $\Delta \eta$ is quite small and 
variations of $\Te(x,y)$ and $\muec(x,y)$ within $\Delta \eta$ are negligible.   
Therefore we calculate the fluxes in the zeroth-order of $\Delta \eta$
(zero edge-width model). 
We also assume that the width of the bulk region $\Delta y$ is small 
compared with the length scale of variations of $\Te$ and $\muec$, 
and calculate the fluxes in the first-order of $\Delta y$ 
(narrow wire model).

\subsection{Edge Currents}
\label{sec:edge_currents}

We introduce coordinates $(\xi, \eta)$ 
for each boundary of the two-dimensional system. 
The unit vector normal to the boundary, directed to the outside 
of the sample, is denoted by $\vn$.  
We take the $\eta$ axis in the direction of $\vn$ 
and the $\xi$ axis along the boundary in the direction of $\hat{\epsilon} \vn$. 

We calculate fluxes in the edge region 
$\eta_{\edge}<\eta<\eta_{\edge}+\Delta \eta$.  
In the region $\eta>\eta_{\edge}+\Delta \eta$ the electron density and the flux densities 
are assumed to be negligible. 
Since $\Delta \eta$ is small in the present steep confining potential,  
the hopping flux in the edge region is negligible. 
The drift flux is not negligible
because of the large gradient of the confining potential. 
With use of eq.(\ref{eq:jn_drift}), the drift number flux is given by 
\begin{equation}
\vJ^{\edge}_n = K_n \hat{\epsilon} \vn  ,
\end{equation}
with
\begin{equation}
K_n=  \frac{s_B}{h} \sum_{\alpha} 
       \int_{\eta_{\edge}}^{\eta_{\edge}+\Delta \eta} d\eta 
      \frac{\partial V}{\partial \eta}
            f( \ve^0_{\alpha}+V, \muec,\Te)   . 
\end{equation}
Since the $\eta$ dependence of $\muec$ and $\Te$ is neglected, 
\begin{equation}
K_n=  \frac{s_B}{h} \sum_{\alpha} 
                 \int_{\ve_{\alpha}}^{\infty}  f( \ve,\muec,\Te) d\ve
	 = - s_B \frac{2\pi l^2}{h} \Omega_0  ,
\end{equation}
where $\ve_{\alpha}$, $\muec$, $\Te$ and $\Omega_0$ 
are to be evaluated at $\eta_{\edge}$.  
The quantitiy $K_n$ is related to the magnetization per unit area, $M$, at $\eta_{\edge}$   
by 
\begin{equation}
K_n(\eta_{\edge})  = (c/e) M(\eta_{\edge}) \ , 
\label{eq:M}
\end{equation}
since the edge electric current $\vI$ and $M$
are related~\cite{Landau1960} by $\vI= - c M  \hat{\epsilon} \vn$. 
The thermal flux is given by
\begin{equation}
\vJ^{\edge}_q = K_q \hat{\epsilon} \vn ,
\end{equation}
with 
\begin{equation}
K_q =  \frac{s_B}{h} \sum_{\alpha} 
                       \int_{\ve_{\alpha}}^{\infty} 
                       (\ve-\muec )  f( \ve, \muec,\Te) d\ve  ,
\label{eq:Kq}
\end{equation}
which is written as 
\begin{equation}
K_q = s_B \frac{2\pi l^2}{h}
             \int_{-\infty}^{\much} 
             \left( \Omega_0 (\much')+ \Te s_0 (\much') \right)  d \much' .
\end{equation}

Derivatives of $K_n$ and $K_q$ are given by 
\begin{equation}
\frac{\partial K_n}{\partial \much} = \Lnnyx , \ \ 
\frac{\partial K_n}{\partial \Te} =   \frac{\Lnqyx}{\Te}  ,
\end{equation}
\begin{equation}
\frac{\partial K_q}{\partial \much} = \K21yx , \ \ 
\frac{\partial K_q}{\partial \Te} =   \frac{\Lqqyx}{\Te}  ,
\label{eq:Lqqyx}
\end{equation}
with
\begin{equation}
\Lnqyx =  s_B \frac{2\pi l^2}{h} \Te s_0 = \K21yx + K_n \ ,
\label{eq:Lnqyx}
\end{equation}
\begin{equation}
\Lqqyx =  s_B \frac{2\pi l^2}{h} \Te^2
             \int_{-\infty}^{\much} 
             \frac{\partial s_0 (\much') }{\partial \Te} d \much' .
\end{equation}
$\Lqqyx$ is also expressed from eqs.(\ref{eq:Kq}) and (\ref{eq:Lqqyx}) by 
\begin{equation}
\Lqqyx = 2 K_q + 
\frac{s_B}{h} \sum_{\alpha} (\ve_{\alpha} -\muec)^2 f_{\alpha} .
\end{equation}

\subsection{Measured Fluxes}

\subsubsection{Fluxes through the Sample}

First we consider fluxes through a cross section of the sample  
along the $x$ direction. 
Each of the number and thermal fluxes is written 
as the sum of the bulk and edge contributions. 

The number flux, denoted as $J^{\tr}_{nx}$, is given by
\begin{equation}
J^{\tr}_{nx}= \int_{0}^{\Delta y} j_{nx} dy 
           + J^{\edge(\Delta y)}_{nx} + J^{\edge(0)}_{nx}  .
\end{equation}
Here $J^{\edge(0)}_{nx}$ and $J^{\edge(\Delta y)}_{nx}$ are 
the edge current flowing in $y<0$ and that in $y>\Delta y$, respectively. 
The sum of these edge currents is zero in equilibrium. 
In nonequilibrium states it is induced 
by the differences 
$\Delta \much= \much (\Delta y) -  \much (0) $ 
and $\Delta \Te= \Te (\Delta y) -  \Te (0) $, 
and is written in their first order as 
\begin{eqnarray}
&J^{\edge(\Delta y)}_{nx} + J^{\edge(0)}_{nx} =  K_n(\Delta y) -K_n(0) \nonumber \\
&=  \Lnnyx  \Delta \much    + \frac{\Lnqyx}{\Te } \Delta \Te   \ .
\end{eqnarray}
The flux density in the bulk region, $j_{nx}$, is given by eq.(\ref{eq:jnx}) 
and is independent of $y$ in the lowest order of $\Delta y$.   
Therefore $J^{\tr}_{nx}$ is given by
\begin{equation}
\frac{J^{\tr}_{nx}}{\Delta y}\!=\! - \Lnnxx  \nabla_x \muec 
                             +\! \Lnnyx \nabla_y \muec
                             -\! \frac{\Lnqxx}{\Te } \nabla_x \Te
                             +\! \frac{\Lnqyx}{\Te } \nabla_y \Te .
\label{eq:jtr_nx}
\end{equation}
The measured number flux $J^{\tr}_{nx}$ is written only 
in terms of $\vnabla \muec$ and $\vnabla \Te$  
since it is zero in equilibrium. 

The thermal flux $J^{\tr}_{qx}$ is given by
\begin{equation}
J^{\tr}_{qx}= \int_{0}^{\Delta y} j_{qx} dy 
           + J^{\edge(\Delta y)\tr}_{qx} + J^{\edge(0)\tr}_{qx}  ,
\end{equation} 
where
\begin{equation}
J^{\edge,\tr}_{qx} =  J^{\edge}_{\epsilon x} - \muec^0 J^{\edge}_{nx}
=  J^{\edge}_{qx} +  (\muec - \muec^0) J^{\edge}_{nx} .
\end{equation} 
Here $\muec^0$ is a reference electrochemical potential  
and lies between $\muec(0)$ and $\muec(\Delta y)$. 
It is the value of the uniform electrochemical potential 
in a fictitous thin electrode, 
which is inserted into the sample 
to measure the fluxes through a cross section at any $x$.   
In the first order of $\Delta y$, the contribution from edges becomes
\begin{equation}
J^{\edge (\Delta y)\tr}_{qx} \!+\! J^{\edge (0)\tr}_{qx} =  \K21yx \Delta \much 
   + \frac{\Lqqyx}{\Te} \Delta \Te      + K_n \Delta \muec  \ ,
\end{equation}
where $\Delta \muec =\muec (\Delta y) -  \muec (0)$. 
The contribution of the last term, $ K_n \Delta \muec$, 
due to the diamagnetic surface current 
has been first pointed out by Obraztsov.~\cite{Obraztsov1964}
Then $J^{\tr}_{qx}$ is given by
\begin{equation}
\frac{J^{\tr}_{qx}}{\Delta y}\!=\! - \Lnqxx \nabla_x \muec 
                             +\!  \Lnqyx \nabla_y \muec
                             -\!  \frac{\Lqqxx}{\Te } \nabla_x \Te 
                             +\! \frac{\Lqqyx}{\Te} \nabla_y \Te \ .
\label{eq:jtr_qx}
\end{equation}

\subsubsection{Fluxes through the Boundary}

Next we consider fluxes along the $y$ direction 
through each of boundaries at $y=0$ and $y=\Delta y$ 
into a fictitious electrode outside the sample.

First we use again the coordinates $(\xi, \eta)$ 
introduced in \S \ref{sec:edge_currents} and 
write fluxes through a segment of the boundary with length $\Delta \xi$ 
in terms of the bulk and edge currents.  
We integrate eq.(\ref{eq:nt}) 
over an infinitesimal region with width $\Delta \xi$ 
in $\eta_{\edge}<\eta<\eta_{\edge}+\Delta \eta$.  
The integration of the left hand side of eq.(\ref{eq:nt})  
is of the order of $\Delta \eta$ and is neglected.  
By considering additionally the number flux through the segment of the boundary, 
which is denoted as $\Delta J^{\tr}_{n\eta}$, 
we have 
\begin{equation}
j_{n\eta}(\eta_{\edge}) \Delta \xi  
 + \left[ J^{\edge}_{n\xi}(\xi) - J^{\edge}_{n\xi}(\xi+\Delta \xi)  \right] 
 = \Delta J^{\tr}_{n\eta}  .  
\end{equation}
By taking the limit of small $\Delta \xi$, we obtain 
\begin{equation}
\frac{\Delta J^{\tr}_{n\eta}}{\Delta \xi} =    j_{n\eta}(\eta_{\edge})
                     -  \nabla_\xi J^{\edge}_{n\xi}  \ .
\label{eq:BCn}
\end{equation}
For the thermal flux, we have from eq.(\ref{eq:tst}) 
\begin{equation}
\frac{\Delta J^{\tr}_{q\eta}}{\Delta \xi}  =    j_{q\eta}(\eta_{\edge})
     -  \nabla_\xi J^{\edge}_{q\xi} 
     -  J^{\edge}_{n\xi} \nabla_\xi \muec  \ .
\label{eq:BCq}
\end{equation}

The above formulas are now applied to the boundaries of the sample 
at $y=0$ and $\Delta y$.  
At $y=\Delta y$, 
\begin{equation}
\nabla_x J^{\edge}_{nx} =\nabla_x K_n
=  \Lnnyx  \nabla_x \much    + \frac{\Lnqyx}{\Te } \nabla_x \Te   \ , 
\end{equation}
and the measured number flux $\Delta J^{\tr}_{ny}$ is obtained to be 
\begin{equation}
\! \frac{\Delta J^{\tr}_{ny}}{\Delta x}\!=\!  - \Lnnyx \nabla_{\!x} \muec
                                                                 \!-\! \Lnnxx  \nabla_{\!y} \muec 
                          \!-\! \frac{\Lnqyx}{\Te } \nabla_{\!x} \Te 
                          \!-\! \frac{\Lnqxx}{\Te } \nabla_{\!y} \Te \ .
\label{eq:dJny}
\end{equation}
The same formula is obtained for the boundary at $y=0$. 
For the thermal flux at $y=\Delta y$, 
\begin{equation}
\nabla_x J^{\edge}_{qx}
=  \K21yx  \nabla_x \much    + \frac{\Lqqyx}{\Te} \nabla_x \Te   \ , 
\end{equation}
and 
the measured thermal flux $\Delta J^{\tr}_{qy}$ is
\begin{equation}
\! \frac{\Delta J^{\tr}_{qy}}{\Delta x}  \!=\!    - \Lnqyx \nabla_{\!x} \muec
                                                                    \!-\! \Lnqxx  \nabla_{\!y} \muec 
                                   \!-\! \frac{\Lqqyx}{\Te} \nabla_{\!x} \Te 
                                   \!-\! \frac{\Lqqxx}{\Te } \nabla_{\!y} \Te\ ,
\label{eq:dJqy}
\end{equation}
which is the case both at $y=0$ and at $y=\Delta y$.

\section{Formulation with Transport Fluxes}

\subsection{Transport Flux Densities}

In this section we introduce a transport flux density at each point within the large sample 
by using the formula for each of the measured fluxes in the narrow system 
in the previous section. 
The transport flux densities, denoted as $\vj_n^{\rm tr}(\vr,t)$ and $\vj_q^{\rm tr}(\vr,t)$, 
are defined at each point $\vr$ by
\begin{equation}
\left( \!\! \begin{array}{c}
                            \vj^{\tr}_{n}(\vr,t) \\  \\
                             \vj^{\tr}_{q}(\vr,t) 
            \end{array}  \!\! \right)
 \!=\!      
\left( \!\! \begin{array}{cc}
                 L^{11}   & \!\!  L^{12}  \\  \\
                 L^{21}   &  L^{22}     
           \end{array}  \!\! \right)   \!\! 
\left(  \!\! \begin{array}{c}
                           -\vnabla \muec           \\  \\
                           -\Te^{-1} \vnabla \Te   
            \end{array}    \!\! \right)  \!, 
\label{eq:transport_flux}
\end{equation}
with
\begin{eqnarray}
L^{11} \!= \!
\left( \!\! \begin{array}{rr}
                           \Lnnxx     &    \Lnnxy    \\  \\
                           \Lnnyx     &    \Lnnyy    
           \end{array}  \!\! \right)   \! ,  \
L^{12} \!= \!
\left( \!\! \begin{array}{rr}
                            \Lnqxx     &     \Lnqxy         \\  \\
                            \Lnqyx     &     \Lnqyy   
           \end{array}  \!\! \right)   \! , 
           \nonumber \\
L^{21} \!= \!
\left( \!\! \begin{array}{rr}
                            \Lqnxx     &     \Lqnxy         \\  \\
                            \Lqnyx     &     \Lqnyy   
           \end{array}  \!\! \right)   \! ,  \ 
L^{22} \!= \!
\left( \!\! \begin{array}{rr}
                            \Lqqxx     &      \Lqqxy     \\  \\
                            \Lqqyx     &      \Lqqyy
           \end{array} \!\! \right)  \! ,
\label{eq:transport_coeff}
\end{eqnarray}
where $\Lqnyx=\Lnqyx$ and,  
since the system is isotropic in the $xy$ plane, 
we have the following symmetry relation: 
$L_{yy}^{ij}=L_{xx}^{ij}$ and $L_{xy}^{ij}=-L_{yx}^{ij}$ 
with $i=1,2$ and $j=1,2$.
The right hand side of eq.(\ref{eq:transport_flux}) is equivalent to 
that of the formulas of the measured fluxes in the previous section. 
We note that the coefficients here depend in general on $x$, $y$, and $t$  
through $\Te (\vr,t)$ and $\much (\vr,t)$.  

Diagonal and off-diagonal transport coefficients have different microscopic origins: 
diagonal transport coefficients are due to hopping processes, 
while off-diagonal ones are due to drift processes. 
Note also that off-diagonal transport coefficients 
are written in terms of thermodynamic quantities 
and do not depend on any transition rates. 

It is straightforward to show from 
eqs. (\ref{eq:jnx}), (\ref{eq:jny}), (\ref{eq:jqx}), (\ref{eq:jqy}),  
and (\ref{eq:transport_flux})
that 
the transport flux densities and the total flux densities are related by 
\begin{equation}
\vj_n = \vj_n^{\rm tr} + \hat{\epsilon} \vnabla M_n    \ , \ \  
\vj_{\epsilon} = \vj_{\epsilon}^{\rm tr} + \hat{\epsilon} \vnabla M_{\epsilon}  \ ,  
\label{eq:total-trans}
\end{equation}
with
\begin{equation}
M_n = - K_n    \ , \ \  
M_{\epsilon} = -(K_q + \muec K_n)  \ ,  
\end{equation}
where $\vj_{\epsilon}^{\tr}$ and $\vj_q^{\tr}$ are related by 
$\vj_q^{\tr}=\vj_{\epsilon}^{\tr} - \muec \vj_n^{\tr}$, and 
$K_n$ and $K_q$ are now to be evaluated at a point $\vr$ within the sample. 
The quantity $M_n$ in the above equation is proportional to the magnetization $M$,   
$M_n=-(c/e)M$, from eq.(\ref{eq:M}). 
The above equation states that 
each of the total flux densities, $\vj_n$ and $\vj_{\epsilon}$, 
is the sum of the transport flux density and 
the current density due to spatial variations of the magnetization,   
which is called the magnetization current density.

\subsection{Equations with Transport Flux Densities}

We now rewrite 
the equations of conservation, eqs.(\ref{eq:nt}) and (\ref{eq:tst}), 
in terms of the transport flux densities. 
Because the divergence of each magnetization current density 
in eq.(\ref{eq:total-trans}) is zero, 
the equations
can be written in terms of $\vj_n^{\tr}$ and 
$\vj_q^{\tr}$ only:  
\begin{equation}
\frac{\partial n}{\partial t} = -\vnabla \cdot \vj_n^{\tr} \ ,  
\label{eq:nt2}
\end{equation}
\begin{equation}
\Te \frac{\partial s}{\partial t} = -\vnabla \cdot \vj_q^{\tr} \ 
                              - \vnabla \muec \cdot   \vj_n^{\tr}   - \PL .
\label{eq:tst2}
\end{equation}
The term $- \vnabla \muec \cdot   \vj_n^{\tr}$ is of the second order of the current, 
while the corresponding term in eq.(\ref{eq:tst}), $- \vnabla \muec \cdot   \vj_n$, 
contains a first-order contribution  
due to the drift component in $\vj_n$ when the confining potential is considered. 

The equation for the time evolution of the entropy density
is obtained from eq.(\ref{eq:tst2}) as
\begin{equation}
\frac{\partial s}{\partial t} = - \vnabla \cdot \vj_s  
      + \left( \frac{ds}{dt} \right)_{\rm i}  
      + \left( \frac{ds}{dt} \right)_{\rm e} \ , 
\label{eq:st}
\end{equation}
where 
\begin{equation}
\vj_s= \Te^{-1} \vj_q^{\tr} , 
\end{equation}
is the entropy flux density, 
\begin{equation}
 \left( \frac{ds}{dt} \right)_{\rm i} =  
      -  \vj_n^{\tr}  \cdot \Te^{-1}  \vnabla \muec      
      -  \vj_q^{\tr} \cdot \Te^{-2} \vnabla \Te  \ , 
\label{eq:entropy_production}
\end{equation}
is the rate of entropy production per unit area at $\vr$,  
and 
\begin{equation}
 \left( \frac{ds}{dt} \right)_{\rm e} =  - \Te^{-1} \PL , 
\end{equation}
is due to the heat transfer to the phonon system.

\subsection{Relations between Transport Coefficients}

The Einstein relation tells that 
the coefficient in front of $\vnabla \mu$ 
and that in front of $\vnabla V$ in each of the flux densities are equal. 
This is satisfied for the transport flux densities 
since they are written in terms of 
$\vnabla \muec = \vnabla \mu  + \vnabla V$. 

The Onsager relation~\cite{Onsager1931a,Onsager1931b} 
tells that 
\begin{equation}
L_{kl}(B)= L_{lk}(-B) ,
\end{equation}
where $L_{kl}(B)$ is the coefficient in the formula of the $k$th flux density
in front of the $l$th thermodynamic force. 
In the present case 
with the entropy production in eq.(\ref{eq:entropy_production}), 
the thermodynamic forces~\cite{Groot1962} 
associated with $\vj_n^{\tr}$ and $\vj_q^{\tr}$ are given by  
\begin{equation}
\vF_n = - \Te^{-1} \vnabla \muec \ , \ \ \vF_q = - \Te^{-2} \vnabla \Te \ ,
\end{equation}
respectively. 
Therefore $L_{kl}(B)$ in this case is equal to 
the corresponding coefficient in eq.(\ref{eq:transport_coeff}) 
except the common factor $\Te^{-1}$. 
The coefficients 
in eq.(\ref{eq:transport_coeff}) have the following symmetry 
with respect to the reversal of $\vB$: 
\begin{eqnarray}
L_{xx}^{ij}(\Te,\much, - B)=L_{xx}^{ij}(\Te,\much,  B) , \\
L_{yx}^{ij}(\Te,\much, - B)= - L_{yx}^{ij}(\Te,\much,  B) ,
\end{eqnarray}
with $i=1,2$ and $j=1,2$. 
Then we confirm that they satisfy the Onsager relation. 

In contrast to the coefficients of the transport flux densities, 
those of the total flux densities 
in eq.(\ref{eq:jnx}), (\ref{eq:jny}), (\ref{eq:jqx}), and (\ref{eq:jqy}) 
do not satisfy the Einstein relation nor the Onsager relation. 
We also note that 
the coefficient of the drift component of the total thermal flux density, $\K21yx$, is 
proportional to $\Te s_0 + \Omega_0$ instead of $\Te s_0$ 
[eq.(\ref{eq:K21})], 
while the corresponding coefficient of the transport flux density, $\Lnqyx $, 
is proportional to $\Te s_0$  [eq.(\ref{eq:Lnqyx})] 
and this transport thermal flux density divided by $\Te$
accords with the definition of the entropy flux density. 

Smr$\check{\rm c}$ka and \Streda\cite{Smrcka1977} have proved that 
the transport coefficients for number and energy fluxes 
in quantizing magnetic fields 
are given by integrals of the zero-temperature conductivity tensor
in the case of elastic scatterings
(the relation for number and thermal fluxes has been written 
in ref.\citen{Streda1984a}). 
The same relation holds in the present case. 
Each coefficient $L_{\alpha \beta}^{ij}$ 
with $i=1,2$, $j=1,2$, $\alpha=x,y$, and $\beta=x,y$ 
can be expressed by
\begin{equation}
L_{\alpha \beta}^{ij} = \int^{\infty}_{-\infty} d\ve 
\left( - \frac{\partial f}{\partial \ve} \right) 
(\ve - \muec)^{i+j-2} L_{\alpha \beta}^{11(0)}(\ve)  ,
\end{equation}
in terms of the zero-temperature coefficients 
\begin{eqnarray}
&L_{xx}^{11(0)}(\ve)= L_{yy}^{11(0)}(\ve) = 
     \sum_{\alpha} D_{\alpha} \delta(\ve- \ve_{\alpha}) , \\
&L_{yx}^{11(0)}(\ve)= -L_{xy}^{11(0)}(\ve) = 
      (s_B/h) \sum_{\alpha} \theta(\ve- \ve_{\alpha})  ,
\end{eqnarray}
where $\theta(\ve)=0$ for $\ve<0$ and $\theta(\ve)=1$ for $\ve>0$.

\section{Linear-Response Ettingshausen Effect}

As an illustration, 
we apply the present theory with the transport fluxes 
to the Ettingshausen effect, 
in which the gradient of the electron temperature is developed 
in the direction perpendicular to the current 
in the presence of the magnetic field. 
A spatial variation of the electron temperature in 
the Ettingshausen effect in a quantum Hall system has been 
studied previously in the nonlinear transport regime.~\cite{Akera2002}
Here we make a linear-response calculation 
and investigate spatial variations and quantum oscillations 
of the electron temperature. 

\subsection{Model and Equations}

We consider a two-dimensional system in the region $-W/2 <y< W/2$. 
We employ the zero edge-width model in \S \ref{sec:model}.  
The following two cases are studied. 
In the first case the width of the sample $W$ is comparable to or larger than 
the length scale of the electron-temperature variation $\lambda$. 
The formula of $\lambda$ is given below. 
In the second case $W$ is much smaller than $\lambda$, 
and here only terms in the lowest order of $W/ \lambda$ are retained. 

We use eqs.(\ref{eq:nt2}) and (\ref{eq:tst2}) 
and restrict the calculation to the linear-response regime and to steady states. 
We also assume a uniformity along $x$ where 
flux densities and thermodynamic quantities have no dependence on $x$. 
The exception is $\muec$ which has a constant gradient along $x$.  
The gradient $\nabla_x \muec$ is also independent of $y$
since $\nabla_y \nabla_x \muec =\nabla_x \nabla_y \muec =0$,  
and is equal to $\nabla_x V=eE_x$.  
Then the equations become
\begin{equation}
\nabla_y j_{ny}^{\tr} =0 \ ,  
\label{eq:n(y)}
\end{equation}
\begin{equation}
\nabla_y j_{qy}^{\tr} + \PL =0 \ . 
\label{eq:q(y)}
\end{equation}
We employ the simplest model of the energy loss $\PL$:  
\begin{equation}
\PL=\Cp [\Te (y) - \TL] \ ,
\end{equation}
where $\Cp$ is a constant.  
The boundary conditions at $y=\pm W/2$ are 
\begin{equation}
j_{ny}^{\tr} =0 \ , \ \  j_{qy}^{\tr} =0 \ ,  
\label{eq:BC}
\end{equation}
since the fluxes to the outside of the sample are absent 
($\Delta J^{\tr}_{ny}=0$ and $\Delta J^{\tr}_{qy}=0$) 
in eqs.(\ref{eq:dJny}) and (\ref{eq:dJqy}). 
Using eq.(\ref{eq:n(y)}) and (\ref{eq:BC}) we have
\begin{equation}
j_{ny}^{\tr} =0 \ , \ \ -W/2 <y< W/2 \ .  
\label{eq:jnytr0}
\end{equation}

\subsection{Spatial Variations}

We substitute $\nabla_x \muec =eE_x$ and $\nabla_x \Te =0$ into 
the formula of $j_{qy}^{\tr}$, 
and use eq.(\ref{eq:jnytr0}) to eliminate $\nabla_y \muec$.  
Then we obtain 
\begin{equation}
j^{\tr}_{qy}=   - \Aqnyx eE_x  - \Aqqyy \nabla_y \Te \  ,
\label{eq:jqytr}
\end{equation}
with
\begin{eqnarray}
&\Aqnyx= \Lqnyx - \Lqnyy (\Lnnyy)^{-1} \Lnnyx  , \\ 
&\Aqqyy= [\Lqqyy - \Lqnyy (\Lnnyy)^{-1} \Lnqyy ] \TL^{-1} .
\end{eqnarray}
The transport coefficients are constant in the bulk region in the linear-response regime. 
By substituting eq.(\ref{eq:jqytr}) into eq.(\ref{eq:q(y)}), 
we obtain the equation for $\Te(y)$: 
\begin{equation}
\Aqqyy  \nabla_y^2  \Te = \Cp (\Te-\TL)  \ , 
\end{equation}
with the boundary condition: 
\begin{equation}
j^{\tr}_{qy}=  - \Aqnyx eE_x  - \Aqqyy \nabla_y \Te  =0, \ y=\pm W/2 \ .
\end{equation}
This boundary condition immediately shows that 
the electric field $E_x$ along the current induces 
the temperature gradient $\nabla_y \Te$. 
Solving the equation, we find the spatial variation of $\Te$ to be  
\begin{equation}
\Te(y) -\TL 
=  T_0 [ e^{-(y+W/2)/\lambda} - e^{(y-W/2)/\lambda}  ] .\  
\label{eq:Te(y)}
\end{equation}
The relaxation length of the $\Te$ deviation is found to be
\begin{equation}
\lambda =  (\Aqqyy/ \Cp)^{1/2}\  , 
\end{equation}
where $\Aqqyy>0$ as is proved from the Schwarz inequality. 
The magnitude of the $\Te$ deviation is 
\begin{equation}
T_0  = (\lambda \Aqnyx / \Aqqyy) (1+e^{-W/\lambda})^{-1} eE_x  \ .
\end{equation}
If we assume that $D_{\alpha}= (2N+1) D_0$, 
$T_0/E_x$ depends on $D_0$ and also on $\Cp$. 

\subsection{Quantum Oscillations}

In order to obtain a universal result independent of $D_0$ and $\Cp$, 
we consider $\nabla_y \Te / \nabla_y \muec$ in the limit of $W \ll \lambda$.
In this limit the $y$ dependence of $\nabla_y \Te$ and $\nabla_y \muec$ 
are negligible from eqs.(\ref{eq:Te(y)}) and (\ref{eq:jnytr0}), respectively. 
We use $\nabla_x \Te =0$ again and 
eliminate $\nabla_x \muec$ by using $j_{ny}^{\tr} =0$.  
We then obtain 
\begin{equation}
j^{\tr}_{qy}=   - \Bqnyy \nabla_y \muec  - \Bqqyy \nabla_y \Te \  .
\end{equation}
with
\begin{eqnarray}
&\Bqnyy= \Lqnyy - \Lqnyx (\Lnnyx)^{-1} \Lnnyy  , \\ 
&\Bqqyy= [\Lqqyy - \Lqnyx (\Lnnyx)^{-1} \Lnqyy ] \TL^{-1} .
\end{eqnarray}
From the boundary condition $j^{\tr}_{qy}=0$, we find
\begin{equation}
\nabla_y \Te / \nabla_y \muec =   - \Bqnyy / \Bqqyy  \  .
\end{equation}
Here we assume that $D_{\alpha}= (2N+1) D_0$. 
Then we find that $\nabla_y \Te / \nabla_y \muec$ is independent of $D_0$ and $\Cp$. 
The dimensionless quantity $\kB \nabla_y \Te  / \nabla_y \muec $ 
is a universal function of $\much/ \hbar \oc$ and $\kB \TL/\hbar \oc$, which is plotted in Fig.1. 
Here the spin splitting is neglected. 
We find that the sign of $\nabla_y \Te$ exhibits quantum oscillations 
as a function of $\much$ at $\kB \TL < \hbar \oc$. 
Detailed analysis of the Ettingshausen effect will be given elsewhere. 
\begin{figure}
\begin{center}
\includegraphics[width=8cm]{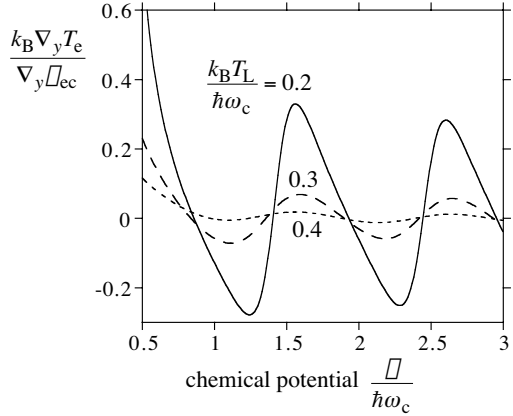}
\end{center}
\caption{
Electron temperature gradient $\nabla_y \Te$ perpendicular to the current  
divided by the Hall field $\nabla_y \muec /e$ ($\muec$: the electrochemical potential)  
is plotted as a function of 
the chemical potential $\much$ for several values of the lattice temperature $\TL$. 
} 
\end{figure}

\section{Discussion}

\subsection{Thermoelectric-Transport Theory}

Nearly half a century ago, 
the thermoelectric coefficients in a magnetic field 
have been calculated quantum-mechanically 
(the references are cited in some papers~\cite{Obraztsov1964,Obraztsov1965,Herring1966}).
Several authors encountered a difficulty 
of violating the Einstein relation 
and the Onsager relation
when calculating the coefficients for the local current density.  
This problem was due to the confusion of the local current density 
with the averaged current density, 
which is defined as 
the measured current through a sample divided by a cross section of the sample.
The Einstein relation and the Onsager relation should be applied to 
the measured current, or to its averaged density, 
and the local current density is different from the averaged current density in a magnetic field 
due to the presence of the diamagnetic surface current 
(a description is given in a textbook~\cite{Landau1960} and a review paper~\cite{Herring1966}). 
Obraztsov~\cite{Obraztsov1964,Obraztsov1965} 
has shown explicitly that the Onsager relation holds 
for the averaged current density 
by calculating the total current through the sample 
including the diamagnetic surface current 
in the case of three-dimensional electron systems. 
He has also shown~\cite{Obraztsov1965} that 
the thermoelectric coefficients in the absence of disorder 
can be expressed in terms of the entropy. 

A current density, 
which satisfies both the Einstein relation and the Onsager relation, 
can be defined also at each point in a sample. 
It has been defined by 
$\vj_{\rm cond}=\vj - c \vnabla \! \times \! \vM$ 
and has been called the conduction current density,~\cite{Landau1960,Herring1966} 
the macroscopic current density,~\cite{Obraztsov1964}
or the transport current density.~\cite{Cooper1997} 
Here $\vj$ is the local current density and 
$c \vnabla \! \times \! \vM$ is the magnetization current density 
with the magnetization $\vM$. 
The current density $\vj_{\rm cond}$ is zero in equilibrium by definition. 

Later the discovery of the quantum Hall effect~\cite{Klitzing1980,Kawaji1981} 
has stimulated theoretical~\cite{Zelenin1982,Girvin1982,Jonson1984,
Streda1983,Streda1984b,Oji1984a,Oji1984b,Oji1985,Zawadzki1984,Grunwald1987,Cooper1997} 
and experimental~\cite{Obloh1984,Gallagher1992} 
studies on thermoelectric effects 
in two-dimensional systems in strong magnetic fields.
The edge current in a two-dimensional system corresponds to 
the surface current in a three-dimensional system. 
The thermoelectric power has been calculated 
in a two-dimensional system of noninteracting electrons 
in a quantizing magnetic field in the absence of disorder~\cite{Zelenin1982,Girvin1982} 
and in the presence of disorder~\cite{Jonson1984,Oji1984b} 
for short-range scatterings
in the self-consistent Born approximation.~\cite{Ando1974} 
Effects of the localization of wave function on the thermoelectric transport 
have been studied~\cite{Grunwald1987} 
within the classical-drift model.~\cite{Iordansky1982,Prange1982,Apenko1985} 
Effects of the interaction between electrons on the thermoelectric transport 
have also been studied.~\cite{Cooper1997}

In this paper 
we have defined a number transport flux density and an energy transport flux density 
from the corresponding measured fluxes in a narrow wire 
and shown that each transport flux density is equal to 
the total flux density minus the magnetization current density, 
namely that corresponding to $\vj - c \vnabla \! \times \! \vM$. 
Note that the current in this paper is due to only the orbital-center motion 
as in refs.\citen{Obraztsov1965,Girvin1982,Grunwald1987}, 
while those in refs.\citen{Landau1960,Herring1966,Obraztsov1964,Cooper1997} 
are due to both the orbital-center and relative motions. 
The localization has been taken into account 
within the classical-drift model,~\cite{Iordansky1982,Prange1982,Apenko1985} 
and the transport coefficients obtained in the present paper 
eq.(\ref{eq:transport_coeff}) reduces to 
those obtained by Grunwald and Hajdu~\cite{Grunwald1987} 
in the absence of hopping processes. 
The transport coefficients in the present paper are 
generalized so that they are applicable 
to a nonlinear transport regime within the local equilibrium 
by including their dependences 
on the local electron temperature and the local chemical potential. 

\subsection
{Hot-Electron Theory and Thermohydrodynamics}

One of the most typical nonlinear effects is the electron heating.  
Uchimura and Uemura~\cite{Uchimura1979} have applied 
the hot-electron theory and the self-consistent Born approximation~\cite{Ando1974} 
to explain the electric-field dependence of the diagonal conductivity, 
observed by Kawaji and Wakabayashi,~\cite{Kawaji1976}
in two-dimensional systems in quantizing magnetic fields.  

Shortly after the discovery of the quantum Hall effect, 
the breakdown of the quantum Hall effect has been 
found,~\cite{Ebert1983,Cage1983,Kuchar1984}  in which 
$\rxx$ increases by several orders of magnitude
when the current is increased up to a critical value.  
As a mechanism of the breakdown, 
a hot-electron model,~\cite{Ebert1983,Gurevich1984,Komiyama1985}  
has been proposed, which combines  
the electron heating and the strong $\Te$ dependence of $\rxx$. 
Later the hot-electron model has been supported by 
the observed spatial evolutions of $\rxx$.~\cite{Komiyama1996,Kaya1998,Kaya1999} 

Gurevich and Mints~\cite{Gurevich1984} have proposed 
a hydrodynamic equation based on the hot-electron model 
to calculate spatio-temporal variations of $\Te$ in quantum Hall systems. 
Their equation, however, is restricted to one-dimensional variations of $\Te$, 
and the thermal flux due to the drift motion is missing. 

The previous theory by one of the present authors~\cite{Akera2002}
has proposed a set of hydrodynamic equations for 
two-dimensional variations of $\Te$ and $\muec$ 
as well as a model of flux densities due to drift and hopping processes,   
which is a generalization of the theory for one-dimensional variations 
by the same author.~\cite{Akera2001} 
In this theory,~\cite{Akera2002} however, 
the drift thermal flux density was not correct: 
the drift thermal flux density in eq.(3.13) there 
was $\vj_q^{\rm drift} - \vj^{\rm drift}_{q0}$ 
where $\vj^{\rm drift}_{q0}$ is the drift thermal flux density at $\Te =0$, 
but it should be $\vj_q^{\rm drift}$. 
Equation (3.13) in ref.\citen{Akera2002}
should be replaced by eq.(\ref{eq:tst}) in the present paper and 
boundary conditions eq.(3.14) should be corrected to
$\vj_n^{\tr} \cdot \vn_{\rm b}=0$, $\vj_q^{\tr} \cdot \vn_{\rm b}=0$ 
where $\vn_{\rm b}$ is the unit vector perpendicular to the boundary. 

\section{Conclusions}

We have developed a theory of thermohydrodynamics in quantum Hall systems  
to study spatio-temporal variations in the linear- and nonlinear-transport regime
in the local-equilibrium approximation.
A nonlinear effect has been taken into account through 
dependences of the transport coefficients 
on the local electron temperature and the local chemical potential.
This theory can be used to investigate, for example,  
the electron temperature distribution 
in the vicinity of the breakdown of the quantum Hall effect 
in a system with compressible and incompressible strips. 

Quantum Hall systems possess several unique features. 
The formation of the Landau levels gives rise to 
quantum oscillations of thermoelectric properties as a function of the filling factor. 
The activation energy for the transport 
plays an essential role in producing the bistability 
in the breakdown of the quantum Hall effect. 
The large drift current is responsible for 
the observed long relaxation length of 
the electron temperature.~\cite{Komiyama1996,Kaya1998,Kaya1999}  
These features will continue to provide a variety of unique thermohydrodynamic phenomena 
in quantum Hall systems.

\section*{Acknowledgments}

The authors would like to thank 
T. Ise, K. Shimoyama, T. Maeda, T. Nakagawa, and S. Kanamaru 
for valuable discussions. 
This work was supported in part by 
the Grant-in-Aid for Scientific Research (C)
from Japan Society for the Promotion of Science.



\begin{thebibliography}{99}

\bibitem{Groot1962} 
S.R. de Groot and P. Mazur: 
{\it Non-equilibrium Thermodynamics} 
(North-Holland, Amsterdam, 1962).

\bibitem{Scholl2001} 
E. Sch{\"o}ll: 
{\it Nonlinear Spatio-Temporal Dynamics and Chaos in Semiconductors} 
(Cambridge University Press, Cambridge, 2001). 

\bibitem{Klitzing1980} 
K. von Klitzing, G. Dorda and M. Pepper: 
Phys. Rev. Lett. {\bf 45} (1980) 494. 

\bibitem{Kawaji1981} 
S. Kawaji and J. Wakabayashi: 
in {\it Physics in High Magnetic Fields} 
(Springer, Berlin, 1981) p.\ 284.

\bibitem{Chklovskii1992}
D. B. Chklovskii, B. I. Shklovskii, and L. I. Glazman: 
Phys. Rev. B {\bf 46} (1992) 4026. 

\bibitem{Ahlswede2001}
E. Ahlswede, P. Weitz, J. Weis, K. von Klitzing, and K. Eberl: 
Physica B {\bf 298} (2001) 562. 

\bibitem{Guven2003}
K. G\"uven and R. R. Gerhardts: 
Phys. Rev. B {\bf 67} (2003) 115327.

\bibitem{Ebert1983} G. Ebert, K. von Klitzing, K. Ploog and G. Weimann: 
J. Phys. C {\bf 16} (1983) 5441.

\bibitem{Cage1983} M. E. Cage, R. F. Dziuba, B. F. Field, 
E. R. Williams, 
S. M. Girvin, A. C. Gossard, D. C. Tsui and R. J. Wagner: 
Phys. Rev. Lett. {\bf 51} (1983) 1374. 

\bibitem{Kuchar1984} F. Kuchar, G. Bauer, G. Weimann and H. Burkhard:
Surf. Sci. {\bf 142} (1984) 196.

\bibitem{Komiyama1996} 
S. Komiyama, Y. Kawaguchi, T. Osada and Y. Shiraki: 
Phys. Rev. Lett. {\bf 77} (1996) 558. 

\bibitem{Kaya1998} I. I. Kaya, G. Nachtwei, K. von Klitzing and K. Eberl: 
Phys. Rev. B {\bf 58} (1998) R7536. 

\bibitem{Kaya1999} I. I. Kaya, G. Nachtwei, K. von Klitzing and K. Eberl: 
Europhys. Lett. {\bf 46} (1999) 62. 

\bibitem{Ahlers1993} F. J. Ahlers, G. Hein, H. Scherer, L. Bliek, 
H. Nickel, R. L{\"o}sch and W. Schlapp: 
Semicond.\ Sci.\ Technol. {\bf 8} (1993) 2062.

\bibitem{Sagol2001} B.E. Sagol, G. Nachtwei, I. I. Kaya, 
K. von Klitzing and K. Eberl: 
Proc. 25th Int. Conf. Phys. Semicond., eds. N. Miura and T. Ando 
(Springer, Berlin, 2001) p. 959.

\bibitem{Kawano2000} Y. Kawano and S. Komiyama: 
Phys. Rev. B {\bf 61} (2000) 2931. 

\bibitem{Kawano2003} Y. Kawano and S. Komiyama: 
Phys. Rev. B {\bf 68} (2003) 085328.

\bibitem{Akera2002} 
H. Akera: J. Phys. Soc. Jpn. {\bf 71} (2002) 228.

\bibitem{Nixon1990} J. A. Nixon and J. H. Davies:
Phys. Rev. B {\bf 41} (1990) 7929.

\bibitem{Iordansky1982} S. V. Iordansky: 
Solid State Commun. {\bf 43} (1982) 1.

\bibitem{Prange1982} R. E. Prange and R. Joynt: 
Phys. Rev. B {\bf 25} (1982) 2943.

\bibitem{Apenko1985} S. M. Apenko and Yu. E. Lozovik: 
J. Phys. C {\bf 18} (1985) 1197. 

\bibitem{Akera2000} 
H. Akera: J. Phys. Soc. Jpn. {\bf 69} (2000) 3174.

\bibitem{Ise2002}
T. Ise and H. Akera: 
J. Phys. Soc. Jpn. {\bf 71} (2002) 1712.

\bibitem{Cooper1997}
N. R. Cooper, B. I. Halperin and I. M. Ruzin: 
Phys. Rev. B {\bf 55} (1997) 2344.

\bibitem{Ando1974}
T. Ando and Y. Uemura: 
J. Phys. Soc. Jpn. {\bf 36} (1974) 959.

\bibitem{Landau1960} 
L. D. Landau and E. M. Lifshitz: 
{\it Electrodynamics of Continuous Media}
(Pergamon Press, Oxford, 1960) \S 29.

\bibitem{Obraztsov1964}
Yu. N. Obraztsov: Fiz. Tverd. Tela (Leningrad) {\bf 6} (1964) 414  
[Sov. Phys. Solid State {\bf 6} (1964) 331]. 

\bibitem{Onsager1931a}
L. Onsager: Phys. Rev. {\bf 37} (1931) 405.

\bibitem{Onsager1931b}
L. Onsager: Phys. Rev. {\bf 38} (1931) 2265.

\bibitem{Smrcka1977}
L. Smr$\check{\rm c}$ka and P. \Streda: J. Phys. C {\bf 10} (1977) 2153. 

\bibitem{Streda1984a} 
P. St$\check{\rm r}$eda and H. Oji: Phys. Lett. {\bf 102A} (1984) 201. 

\bibitem{Grunwald1987}
A. Grunwald and J. Hajdu: 
Solid State Commun. {\bf 63} (1987) 289. 

\bibitem{Obraztsov1965}
Yu. N. Obraztsov: Fiz. Tverd. Tela (Leningrad) {\bf 7} (1965) 573  
[Sov. Phys. Solid State {\bf 7} (1965) 455]. 

\bibitem{Herring1966}
C. Herring: 
J. Phys. Soc. Jpn. {\bf 21}, Suppl. (1966) v.

\bibitem{Zelenin1982}
S. P. Zelenin, A. S. Kondrat'ev, and A. E. Kuchma: 
Fiz. Tekh. Poluprovodn. {\bf 16} (1982) 551  
[Sov. Phys. Semicond. {\bf 16} (1982) 355]. 

\bibitem{Girvin1982}
S. M. Girvin and M. Jonson: J. Phys. C {\bf 15} (1982) L1147. 

\bibitem{Jonson1984}
M. Jonson and S. M. Girvin: Phys. Rev. B {\bf 29} (1984) 1939.

\bibitem{Streda1983}
P. \Streda: J. Phys. C {\bf 16} (1983) L369. 

\bibitem{Streda1984b}
P. \Streda: Phys. Status Solidi. (b) {\bf 125} (1984) 849. 

\bibitem{Oji1984a}
H. Oji: Phys. Rev. B {\bf 29} (1984) 3148.

\bibitem{Oji1984b}
H. Oji: J. Phys. C {\bf 17} (1984) 3059.

\bibitem{Oji1985}
H. Oji and P. \Streda: Phys. Rev. B {\bf 31} (1985) 7291.

\bibitem{Zawadzki1984}
W. Zawadzki and R. Lassnig: Surf. Sci. {\bf 142} (1984) 225. 

\bibitem{Obloh1984}
H. Obloh, K. von Klitzing, and K. Ploog: 
Surf. Sci. {\bf 142} (1984) 236.

\bibitem{Gallagher1992} 
B. L. Gallagher and P. N. Butcher: 
in {\it Handbook on Semiconductors}, edited by P. T. Landsberg 
(North-Holland, Amsterdam, 1992), Vol. 1, p. 721.

\bibitem{Uchimura1979}
N. Uchimura and Y. Uemura: 
J. Phys. Soc. Jpn. {\bf 47} (1979) 1417.

\bibitem{Kawaji1976}
S. Kawaji and J. Wakabayashi: Surf. Sci. {\bf 58} (1976) 238.

\bibitem{Gurevich1984} A. Vl. Gurevich and R. G. Mints: 
Pis'ma Zh. Eksp. Teor. Fiz. {\bf 39} (1984) 318 
[JETP Lett. {\bf 39} (1984) 381]. 

\bibitem{Komiyama1985} 
S. Komiyama, T. Takamasu, S. Hiyamizu and S. Sasa: 
Solid State Commun. {\bf 54} (1985) 479.

\bibitem{Akera2001} H. Akera: J. Phys. Soc. Jpn. {\bf 70} (2001) 1468.

\end{thebibliography}
\end{document}